\documentclass[12pt]{article}
\usepackage{natbib,graphicx,psfrag,amsmath,amssymb}
\usepackage{times}
\usepackage{graphics}
\newcommand{\el}{\ell_{\rm seg}}
\newcommand{\T}{T_{\rm tot}}
\newcommand{\be}{\begin{equation}}
\newcommand{\ee}{\end{equation}}
\begin{document}
\title{Clustering of SNPs along a chromosome: can the neutral model
be rejected?}
\author{Anders Eriksson$^1$, Bernhard Haubold$^2$, and Bernhard
Mehlig$^1$}
\maketitle

\noindent \mbox{}$^1$Physics \& Engineering Physics, Chalmers/GU, Gothenburg, Sweden\\
\noindent \mbox{}$^2$LION Bioscience AG, Waldhofer Str. 98, 69123 Heidelberg, Germany

\vspace{1cm}

Running Title: SNP Clustering

\begin{abstract}
Single nucleotide polymorphisms (SNPs) often appear in clusters along the length
of a chromosome. This is due to variation in local coalescent times caused by,
for example, selection or recombination. Here we investigate whether
recombination alone (within a neutral model) can cause statistically significant
SNP clustering. We measure the extent of SNP clustering as the ratio between the
variance of SNPs found in bins of length $l$, and the mean number of SNPs in
such bins, $\sigma^2_l/\mu_l$. For a uniform SNP distribution
$\sigma^2_l/\mu_l=1$, for clustered SNPs $\sigma^2_l/\mu_l > 1$. Apart from the
bin length, three length scales are important when accounting for SNP
clustering: The mean distance between neighboring SNPs, $\Delta$, the mean
length of chromosome segments with constant time to the most recent common ancestor,
$\el$, and the total length of the chromosome, $L$. We show 
that SNP clustering is observed if $\Delta < \el
\ll L$. Moreover, if $l\ll \el \ll L$, clustering becomes 
independent of the rate
of recombination.  We apply our results
to the analysis of SNP data sets from mice, and human chromosomes
6 and X. Of the three data sets investigated, the human X chromosome displays
the most significant deviation from neutrality.
\end{abstract}

\begin{figure}
\psfrag{YLABEL}{\large$\lambda(x)$} 
\psfrag{XLABEL}{\large$x/L$}
 \centerline{\includegraphics[clip,width=12cm]{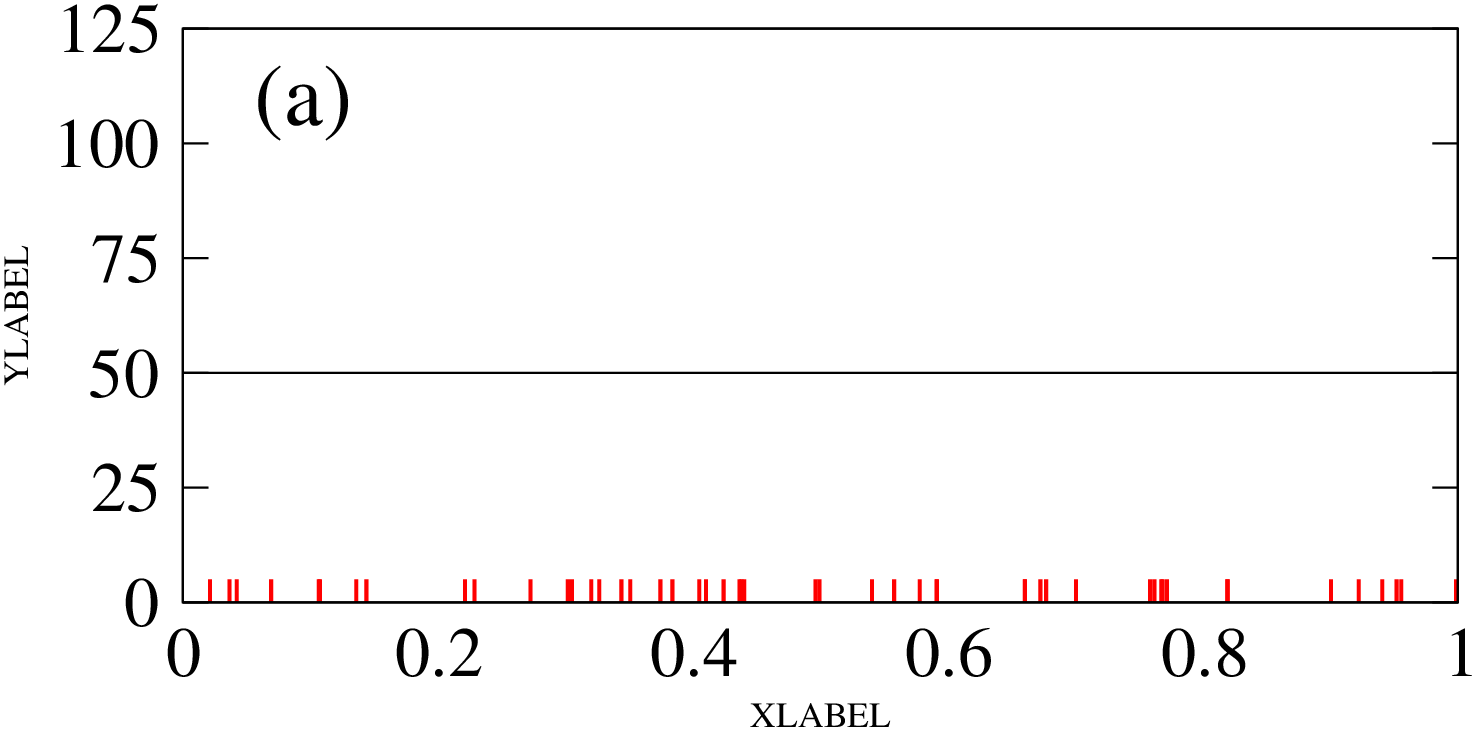}}

 \centerline{\includegraphics[clip,width=12cm]{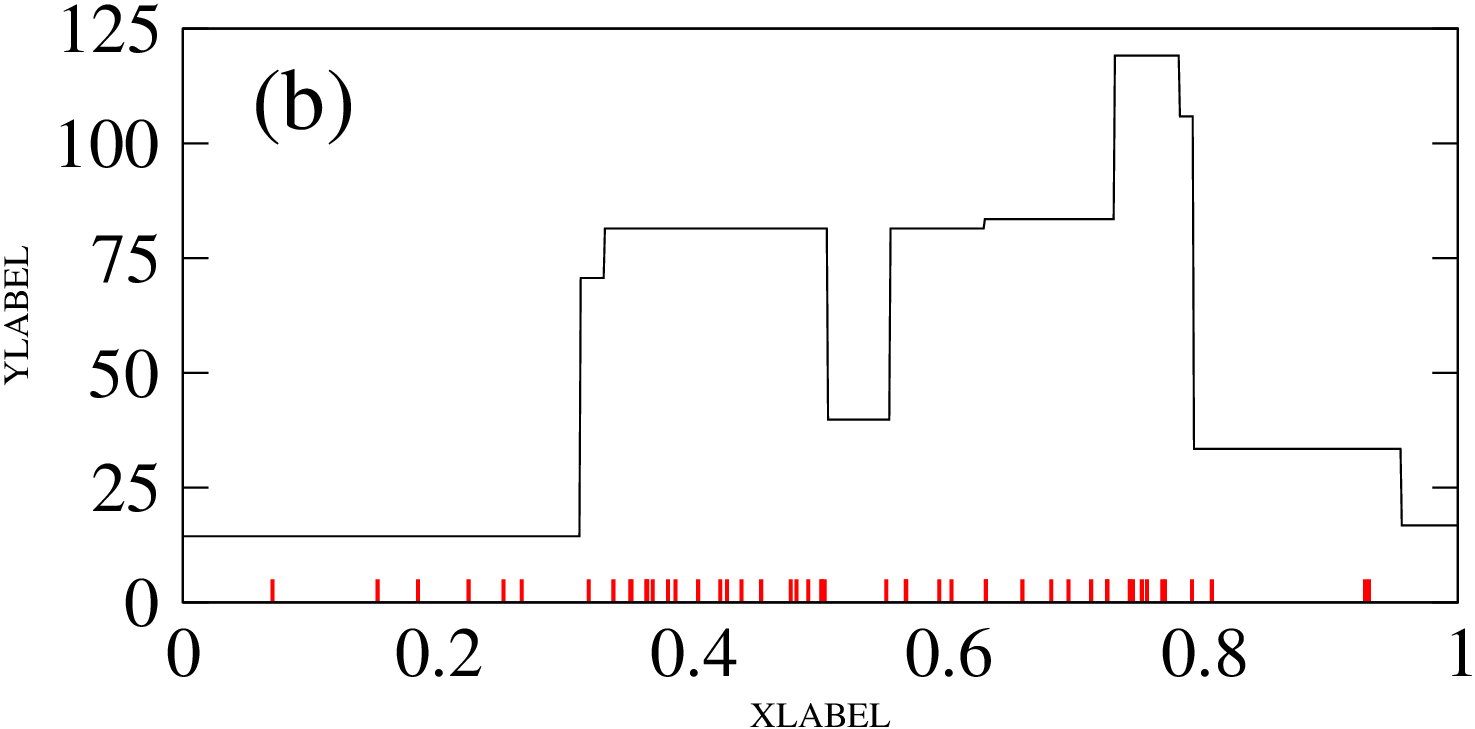}}

 \centerline{\includegraphics[clip,width=12cm]{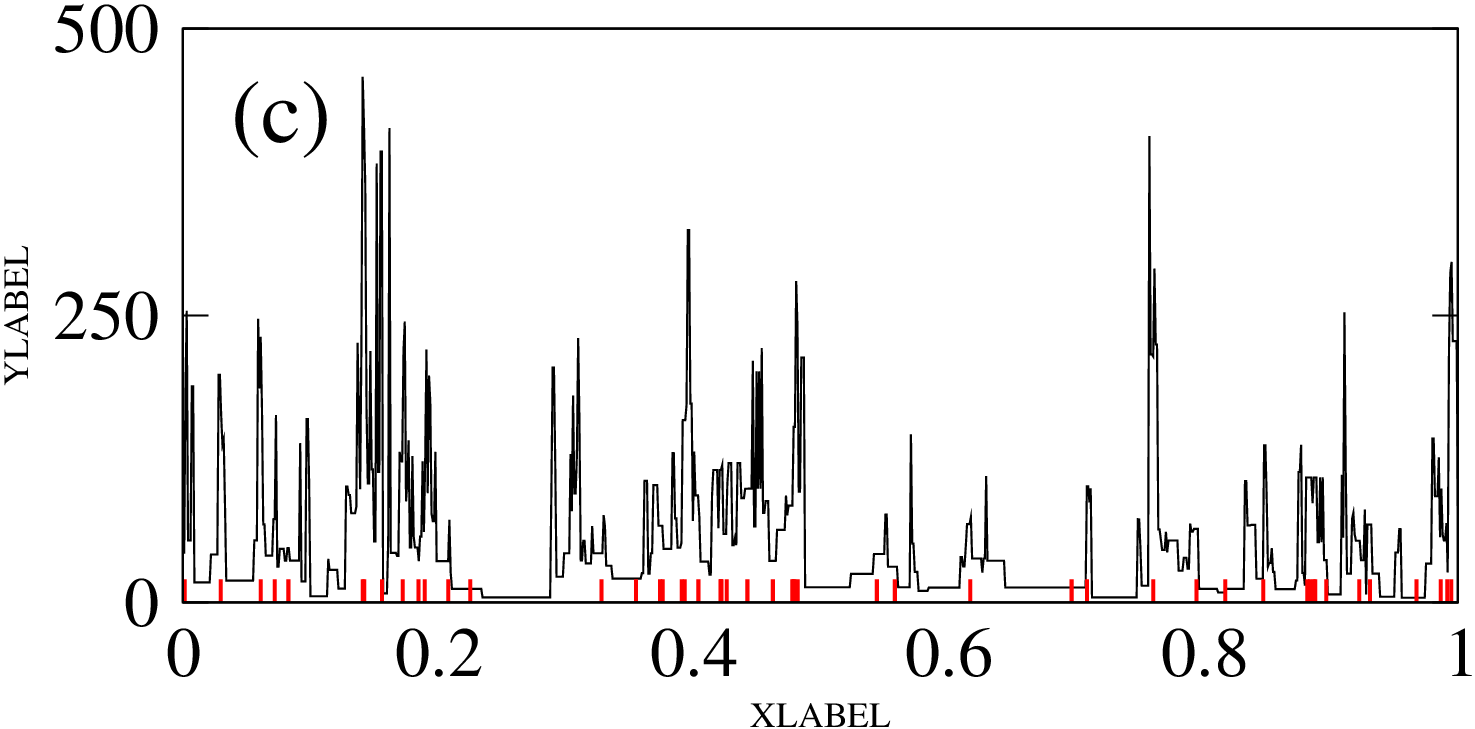}}
\caption{\label{fig0} Recombination leads to spatial
fluctuations in local coalescent
times, $\T(x)$ \citep{hud90:gen}, which in turn cause fluctuations of the local
mutation rate $\lambda(x)$ (solid lines). Shown are three realizations of
$\lambda(x)$ together with the locations of $S=50$ SNPs (vertical bars) for
$n\!=\!2$ and (a) $R=0$, (b) $R=10$, and (c) $R=1000$.}
\end{figure}

\begin{figure}
\psfrag{TAG1}{$l$} 
\psfrag{TAG2}{$\Delta$} 
\psfrag{TAG3}{$\el$} 
\psfrag{TAG4}{$L$} 
\psfrag{N1}{$n_1(l) = 8$} 
\psfrag{N2}{$n_2(l) = 1$} 
\psfrag{N3}{$n_3(l) = 6$} 
 \centerline{\includegraphics[clip,width=12cm]{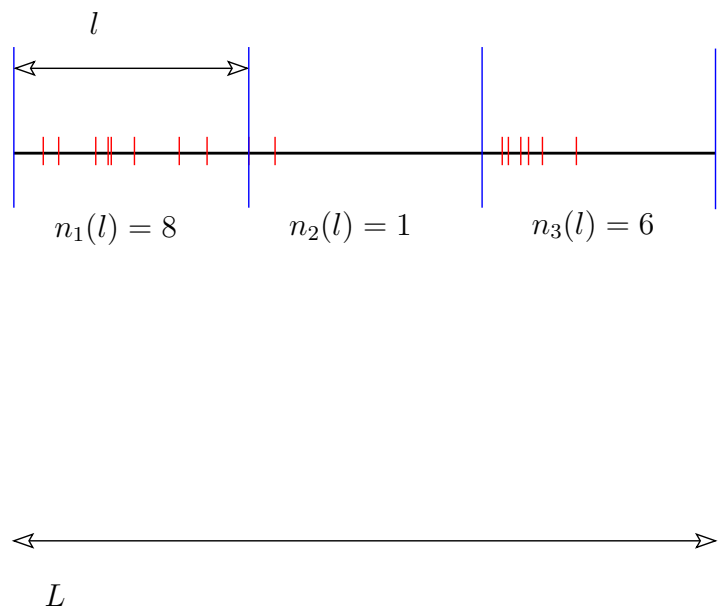}}
 \caption{\label{fig:scales} A chromosome of length $L$ is divided
 into contiguous bins of length $l$. The number
 of SNPs in bin $j$ is denoted by $n_j(l)$. }
\end{figure}

\noindent {\sc Introduction}
\\[0.2cm]
Single nucleotide polymorphisms (SNPs) are the most abundant polymorphisms in
most populations.  Due to their ubiquity and stability they are useful in the
diagnosis of human diseases \citep{zho02:cou}, detection of human disease genes
\citep{wil02:los}, and gene mapping in organisms as diverse as humans
\citep{mci01:fin}, \textit{Arabidopsis thaliana} \citep{cho99:gen}, and
\textit{Drosophila} \citep{ber01:gen}. For
this reason, several large-scale SNP-mapping projects are currently under way in
eukaryotic model organisms including \textit{A. thaliana}
(http://arabidopsis.org/Cereon), \textit{Drosophila} \citep{hos01:sin}, mouse
\citep{lin00:lar}, and human \citep{int01:ini,tsc01:map}.
\\[0.2cm]
A central question in the analysis of data collected in the context of these
projects is how SNPs are distributed along a chromosome and 
what inferences about selection might be drawn from this distribution.
 This question can be addressed at the level of individual polymorphisms
\citep{fay01:pos} or at the level of the whole genome \citep{lin00:lar}. 
\\[0.2cm]
\citet{lin00:lar} have observed that SNPs cluster along chromosomes in
mice.
This clustering may either be due to variation in local
mutation rates, or variation in local coalescent times. 
The hypothesis of local differences in mutation rates in
mice was rejected, leaving differences in local coalescent times as the most
likely explanation of SNP clustering. In the case
of the mouse genome such variation in coalescent times may be due to selection
in the wild or selection for unusual coat colors (c.f. breeding of `fancy' mice
in the eighteenth and nineteenth centuries). Another possibility mentioned  is
the effect of inbreeding \citep{lin00:lar}.
\\[0.2cm]
On the other hand, recombination alone leads to fluctuations in the time to the
most recent common ancestor along a chromosome \citep{hud90:gen}. Since time to the most recent
common ancestor is proportional to the number of SNPs found in the respective
chromosome segment, recombination in a neutral model might be sufficient to
account for genome-wide SNP clustering.
\\[0.2cm]
A well established stochastic model for neutral genetic variation
is the constant-rate mutation coalescent process under the infinite-sites model.
According to this model, the total number of SNPs found in a
sample
is expected to be Poisson distributed with parameter $\lambda = \theta
\T/2$,
where $\T$ is the total time to the most recent common ancestor, $\theta = 4N_{\rm e}u$, $u$ is the probability of mutation per site per generation, and $N_{\rm e}$ is the effective population size
(see \citet{hud90:gen}
for a review).  This is a {\em global} property of any contiguous
stretch of
DNA, and holds in the absence of recombination, where all sites have the
same
genealogy (and thus the total
time $\T$ to the most recent common ancestor is constant
along the chromosome).
In the presence of recombination, the number of polymorphisms
conditional on the genealogies of all sites is still Poisson distributed
with
parameter
\begin{equation}
\frac{\theta}{2} \int_0^L{\rm d}x \,\T(x)\,.
\end{equation}
Here $x$ denotes the position on,
and $L$ the length of the chromosome.
Since the value of parameter (1)
fluctuates between samples, the total number of SNPs is no longer
Poisson
distributed, except in the case of very frequent recombination where
the variance of this parameter tends to zero. These properties
of the coalescent are reviewed in \cite{hud90:gen}. For more recent
reviews see \cite{nor01:coa} and \cite{nor02:lin}.
\\[0.2cm]
In this paper we investigate {\em local} SNP statistics:
local spatial fluctuations in $\T(x)$ due to recombination 
\citep{hud90:gen} may give rise to {\em local} variations in the SNP density.
Here we study the implications of
this idea for the analysis of experimental SNP data.
\\[0.2cm]
Specifically, we address the following five questions. How
significant is SNP clustering caused by recombination? 
How does the clustering
depend on the parameters of the model (the sample size, the mutation rate,
 and the recombination rate)?  On which length scales are
such clusters expected?  How does the clustering depend
on the length scale on which it is observed? 
Finally, can recombination alone account for the
clustering of SNPs observed in mice \citep{lin00:lar}, or in the
human genome \citep{tsc01:map}?  In the following
these questions are answered by analyzing coalescent simulations.
\\[0.6cm]
\noindent {\sc Model and methods} \\[0.2cm]
We use coalescent simulations under the neutral infinite-sites model
to generate allele samples \citep{hud90:gen,nor01:coa}. As usual, this
model
incorporates mutation (with rate $\theta = 4 N_{\rm e} u$)
and reciprocal recombination (with rate $R = 4
N_{\rm e} r$, where $r$ is the probability of a recombination event per generation per
sequence). 
\\[0.2cm]
The coalescent process generates genealogies for
all sites of the $n$ sequences in a given sample.  In the absence of
recombination, these genealogies are identical for all sites $x$. In particular,
the total time to the most recent common ancestor $\T$ is the same for all
sites. For a given genealogy, mutations are generated as a Poisson process with
rate $\lambda = \theta \T/2$. This implies that the density of SNPs along the
genome is uniform: in this case SNPs do not cluster.
\\[0.2cm]
If the recombination rate is non-zero, the total time to the most recent
common ancestor, $\T(x)$, varies as a function of the position $x$
\citep{hud90:gen}.
This corresponds to fluctuating local mutation rates 
$\lambda(x) = \theta\,\T(x)/2$.
In the presence of recombination, the distribution of SNPs is thus
determined by a Poisson process in $x$ with fluctuating rates
$\lambda(x)$.
Figure~\ref{fig0} shows such a process for realizations
of $\lambda(x)$ corresponding to three different sets of parameter values. The
fluctuating rates $\lambda(x)$ are shown as solid lines. Note that $\lambda(x)$
is constant over segments of the chromosome which are identical by descent 
(called MRCA segments in the following).  
The
figure illustrates possible local clustering of SNPs as a consequence of local
variation in $\lambda(x)$ due to recombination. 
While the density of SNPs in the top and bottom
panels is uniform, the middle panel exhibits clustering in regions of
high $\lambda(x)$.
\\[0.2cm]
In the remainder of this paper,
the local clustering such as that exhibited in the middle
panel of Figure~\ref{fig0} is described quantitatively. It is customary in
experimental SNP surveys to count SNPs in bins of length $l$. Such a bin might,
for example, correspond to a sequence tagged site (STS), or some arbitrarily
chosen stretch of sequence. The mean number of SNPs per bin is then
\begin{equation}
\label{eq:sigma}
\mu_l = \frac{1}{N_{\rm bins}} \sum_{j=1}^{N_{\rm bins}} n_j(l)
\end{equation}
and its variance
\begin{equation}
\label{eq:var}
\sigma_l^2 = \frac{1}{N_{\rm bins}-1} \sum_{j=1}^{N_{\rm bins}}
\big(n_j(l)-\mu_l\big)^2,
\end{equation}
where 
$n_j(l)$ is the number of SNPs in bin $j$ and $N_{\rm bins}$ is 
the total number of
bins surveyed (c.f. Figure~\ref{fig:scales}).
In some SNP studies, the bins are arranged contiguously along
the chromosome (as depicted in Figure~\ref{fig:scales}), in some
cases the bins are randomly distributed, or equidistributed
but non-contiguous.
\\[0.2cm]
We compare empirical values for $\sigma_l^2/\mu_l$ 
with results of coalescent simulations. In these simulations
we determine the ensemble average (denoted
by $\langle \sigma_l^2\rangle$ in the following) and
the distribution of $\sigma_l^2$ over random
genealogies with mutations. We keep
the total number of mutations, $S$, fixed to the empirical value.
The local rates $\lambda(x)$ are then given by
\begin{equation}
\label{eq:rate}
\lambda(x) = (S/L)\, \T(x)/2\,,
\end{equation}
and the value of $\mu_l$ is constant between
different realizations of the ensemble. 
One has
$\mu_l = l/\Delta$ where $\Delta \equiv L/S$ is the mean distance between
neighboring SNPs\footnote{If $S$ fluctuates from sample to sample,
$\Delta = L/\langle S\rangle = \left[\theta
\sum_{k=1}^{n-1}k^{-1}\right]^{-1}$. Here $n$ is the sample size.}.
  For uniformly distributed SNPs generated with an $x$-independent
  rate, $\sigma_l^2/\mu_l = 1$. 
 In the case of fluctuating rates
\begin{equation}
 \sigma_l^2/\mu_l > 1 
 \end{equation}
is 
expected, since spatial fluctuations of $\lambda(x)$ give rise to an increased
 ``compressibility'' of the sequence of SNPs.
 In other words, $\sigma_l^2/\mu_l$ 
 measures the ``compressibility'' of the sequence
 of SNPs: the larger $\sigma_l^2/\mu_l$, the more significant
 SNP clustering is on scales $l$ and larger.  
\\[0.2cm]
To meaningfully speak about SNP clustering, it is necessary that $\Delta$ is
much smaller than $L$.  This is the case considered in the following. In
addition, we assume that bins are much shorter than the chromosome on which they
are placed, i.e., we make the following assumptions
\begin{equation}
l\ll L\qquad \mbox{and}\qquad \Delta  \ll L\,.
\end{equation}
It is clear (Figure~\ref{fig0}) that the
statistical properties of $\sigma_l^2/\mu_l$ 
crucially depend on how rapidly the rate $\lambda(x)$
fluctuates as a function of $x$. It is convenient
to define a length scale $\el$ as the ratio
of $L$ and the (average) number of jumps of
$\lambda(x)$ along the total length of the chromosome. 
This length scale
corresponds to the average MRCA segment length
in Figure~\ref{fig0}. Here the average is over
the chromosome for a given realization of $\lambda(x)$
as well as over an ensemble of such realizations;
$\el$ depends on the recombination
rate $R$, the sample size $n$, and $L$ \citep{gri97:anc}.
The relative sizes of the mean spacing $\Delta$
between neighboring SNPs, of the bin size $l$,
the chromosome length $L$, and of the average
MRCA segment length $\el$ 
will play a crucial role in determining SNP clustering.
\\[0.2cm]
In the following section, we analyze local SNP clustering
in the model described above. We determine the significance
of the four length scales $\Delta, l, \el$, and $L$ 
for 
the statistics of the observable $\sigma_l^2/\mu_l$ and analyze
for which parameter values $\theta,R$, and on which length
scales SNP clustering due to recombination is expected
to be most significant.
In the final section, 
we discuss the implications of our results in relation to
genome-wide surveys of SNPs in mice and humans.
\\[0.6cm]
\noindent{\sc Analysis of SNP clustering}\\[-1.cm]
\paragraph{Characterization of the spatial fluctuations of $\lambda(x)$:}
For a given genealogy 
under the neutral infinite-sites model with recombination
SNPs are distributed
according to an inhomogeneous Poisson process, that is,
according to a Poisson process with a rate $\lambda(x)$ varying
along the chromosome (see Figures~\ref{fig0}b and c). Given the function $\lambda(x)$, 
the probability of observing $n(l)=k$ SNPs in bin $[0,l]$ is
\begin{equation}
\label{eq:inh_poisson}
P\big(n(l)=k\big|\lambda(x)\big)=\frac{1}{k!}\Big(\int_0^l\!{\rm
d}x\,\lambda(x)\Big)^k\,\exp\Big(-\int_0^l\!{\rm
d}x\,\lambda(x)\Big)  \,.
\end{equation}
Moreover, given the function $\lambda(x)$,
counts of SNPs in non-overlapping bins are  statistically
independent.
\\[0.2cm]
Theoretical predictions are computed as ensemble
averages over random genealogies, corresponding to
 averages over random functions $\lambda(x)$.
These ensemble averages
introduce correlations in the combined process. 
Such correlations may be weak, but they can be 
long-ranged. Their range is determined by the length scale 
on which the random rate $\lambda(x)$ varies. As Figure~\ref{fig0}
shows, $\lambda(x)$ is a piecewise constant function: 
along an MRCA segment the rate is constant, and varies
between MRCA segments. 
The three panels in Figure~\ref{fig0} 
correspond to the three cases $\el = L$ (a), 
$\Delta \ll \el \ll L$ (b), and $\el \ll \Delta$ (c).  The
average MRCA segment length depends on the sample size $n$ as well as on the
recombination rate $R$.
According to \citet{gri97:anc}
\begin{equation}
\label{eq:ell}
\el = L\,\{1+[1-2R/n/(n+1)]\}^{-1}\,,
\end{equation}
where the denominator denotes the expected number
of changes of ancestor along the chromosome
[notice that eq. (\ref{eq:ell}) does not describe
the expected number of sites with the same MRCA
as pointed out by \cite{wiu99:anc}].
The length $\el$ describes the scale
on which the correlations between local mutation rates $\lambda(x)$ decay. 
For $x < \el$ these correlations are strong. For $x\gg \el$, on the
other hand, the correlation function
\begin{equation}
C(x) = \frac{\langle\lambda(y)\lambda(y+x)\rangle
-\langle\lambda(y)\rangle^2}{\langle\lambda^2(y)\rangle
-\langle\lambda(y)\rangle^2}
\end{equation}
decays to zero.
\citet{kap85:use} have shown that
correlations between the times $\T$
pertaining to two loci in an $m$-locus model         
decay according to $C^{-1}$ for large values of $C$
(here $C$ is the recombination rate between those loci).
Identifying $C$ with an effective recombination rate $R_{\rm eff}  = x R/L$ 
(where $R$ is the recombination rate between the ends of the chromosome) one concludes
that $C(x)$ decays as $x^{-1}$ for large $x$. 
In summary, for any finite value of $R$, correlations between local mutation rates
are large on scales up to $\el$, and decay for
larger distances.  
These correlations may affect the fluctuations of $\sigma_l^2$.

\paragraph{Fluctuations of $\sigma_l^2$:} 
The empirical observable $\sigma_l^2/\mu_l$ is expected to fluctuate
from sample to sample\footnote{In our simulations $\sigma_l^2$ does, but
$\mu_l$ does not, since $S$ and $L$ are constant.}.
Since correlations between $\lambda(y)$ and $\lambda(y+x)$
decay as $|x|$ grows, the fluctuations  of $\sigma_l^2/\mu_l$
 tend to zero in the limit of infinite $L$ (with $\el$ and $l$ constant):
 \begin{equation}
 \label{eq:limit}
\begin{array}{l}
\mbox{lim}\qquad
 \sigma_l^2/\mu_l = \langle \sigma_l^2\rangle/\mu_l\,.
\\[-0.2cm]
\scriptstyle L\rightarrow\infty\\[-0.2cm]
\scriptstyle \el\,,\,l\,\,\, \mbox{\small finite}
\end{array}
 \end{equation}
 In this limit the process is thus self-averaging (ergodic):
 eq.~(\ref{eq:limit}) implies that
 the averages of $n(l)$ and of its moments along the chromosome
 equal the ensemble averages $\langle n(l) \rangle$
 of $n(l)$ (and of its moments), see also \citep{plu96:opt}.
Here $n(l)$ is the SNP count in one bin of a given
sample,
and the ensemble average is taken over
random genealogies with mutations.
For the case where 
$n=2$, $\langle n^2(l)\rangle$
can be derived from
eq. (15) in \citep{hud90:gen} by replacing
the recombination rate in this equation with $l R/L$.
One obtains
\begin{eqnarray}
\label{eq:hudson1}
\sigma_l^2/\mu_l
&\simeq& 1+\frac{l}{\Delta}\frac{2}{C^2}
\Big[ -C + \frac{23 C+101}{2 \sqrt{97}}
 \mbox{log}\Big(\frac{2C +13 -\sqrt{97}}{2C+13+\sqrt{97}}
          \frac{13+\sqrt{97}}{13-\sqrt{97}}\Big)\\
             &+&\frac{C-5}{2}\log\Big(\frac{C^2+13 C + 18}{18}\Big)\Big] 
             \nonumber \end{eqnarray} 
 with $C = l R/L$.  
 Eq. (\ref{eq:hudson1}) 
 is approximate (it was derived for an $m$-site
 model in which each site obeys the infinite-sites assumption,
 in the limit of $m\rightarrow\infty$).

When $L$ is large (much larger than $l$ and $\el$)
but finite, the distribution of $\sigma_l^2/\mu_l$ is expected to be narrow,
since $\sigma_l^2/\mu_l$ itself is an average over a large number of
(approximately) independent bins.  When the sample-to-sample variations of
$\sigma_l^2/\mu_l$ are small, theoretical models for SNP clustering are easily
tested (and possibly rejected). Empirically determined observables [such as, for
example, the variance in the number of loci that differ between pairs of
haplotypes in \citep{hau02:rec}] often have broad distributions; it is thus
significant that in recent years longer and longer contiguous chromosome
segments have been sequenced and locally averaged observables such as
$\sigma_l^2/\mu_l$ are now available.  In empirical studies usually $l\ll L$.
Consequently, it is the ratio $\el/L$ which determines the statistics of
$\sigma_l^2/\mu_l$ .
\\[0.2cm]
As pointed out above, recent empirical data
for $\sigma_l^2$ were obtained for 
contiguous non-overlapping bins [as depicted in Figure~\ref{fig:scales}
\citep{tsc01:map}].
In other cases, however, the bins were
randomly distributed over the chromosome \citep{lin00:lar},
or equally spaced but far apart from each other \citep{tsc01:map}.
How does the statistics of $\sigma_l^2$ depend
on the number and the distribution of bins
over the chromosome?  The expected value of $\sigma_l^2$
is independent of the distribution of bins 
[if $l,\el \ll L$ it is approximately given by eq.~(\ref{eq:hudson1})].  
The fluctuations of $\sigma_l^2$, however, critically
depend on the number and distributions of bins.
The following analysis is performed assuming
contiguous bins. When comparing
with empirical data, however, 
confidence intervals for $\sigma_l^2/\mu_l$
were obtained using the empirical number and distribution
of bins.
\\[0.2cm]
Finally, as $\el$ approaches $L$, the fluctuations
of $\sigma_l^2/\mu_l$ are expected to increase.
In the absence of recombination ($\el = L$), SNPs
are distributed according to a Poisson process
and the fluctuations tend to zero (if $l\ll L$).
\\[0.2cm]

\begin{figure}
\psfrag{YLABEL}{\large$\langle\sigma_l^2\rangle/\mu_l$} 
\psfrag{XLABEL}{\raisebox{-0.2cm}{\large$\el/L$}}
 \centerline{\includegraphics[clip,width=12cm]{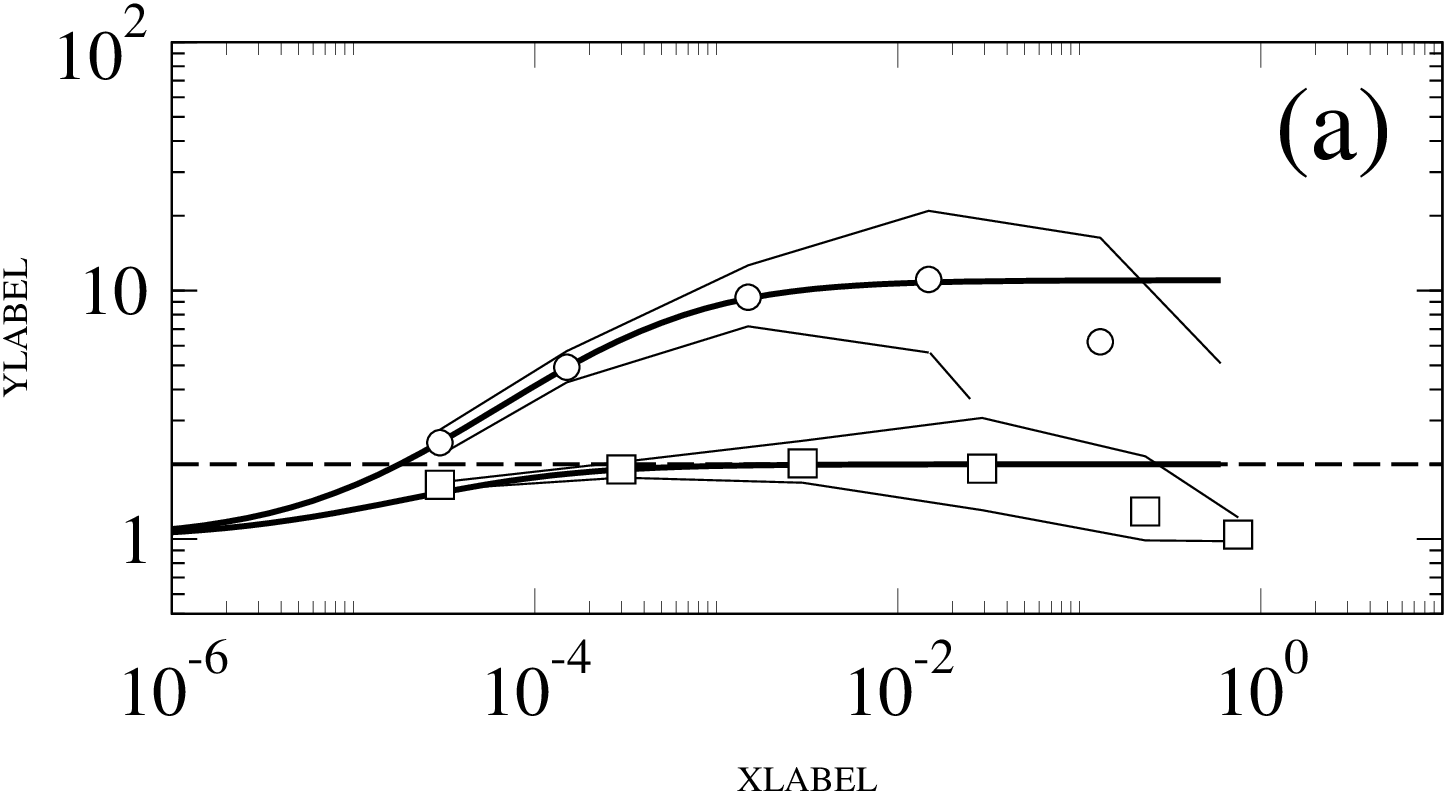}}
\mbox{}\\
\psfrag{YLABEL}{\large$\langle\sigma_l^2\rangle/\mu_l$} 
\psfrag{XLABEL}{\raisebox{-0.2cm}{\large$l/\el$}}
\psfrag{DELTA}{\hspace*{2.5mm}\large$\Delta$} 
\psfrag{TAG1}{$\el/L=6.1\times 10^{-4}$}
\psfrag{TAG2}{$\el/L=1.5 \times 10^{-4}$}
 \centerline{\includegraphics[clip,width=12cm]{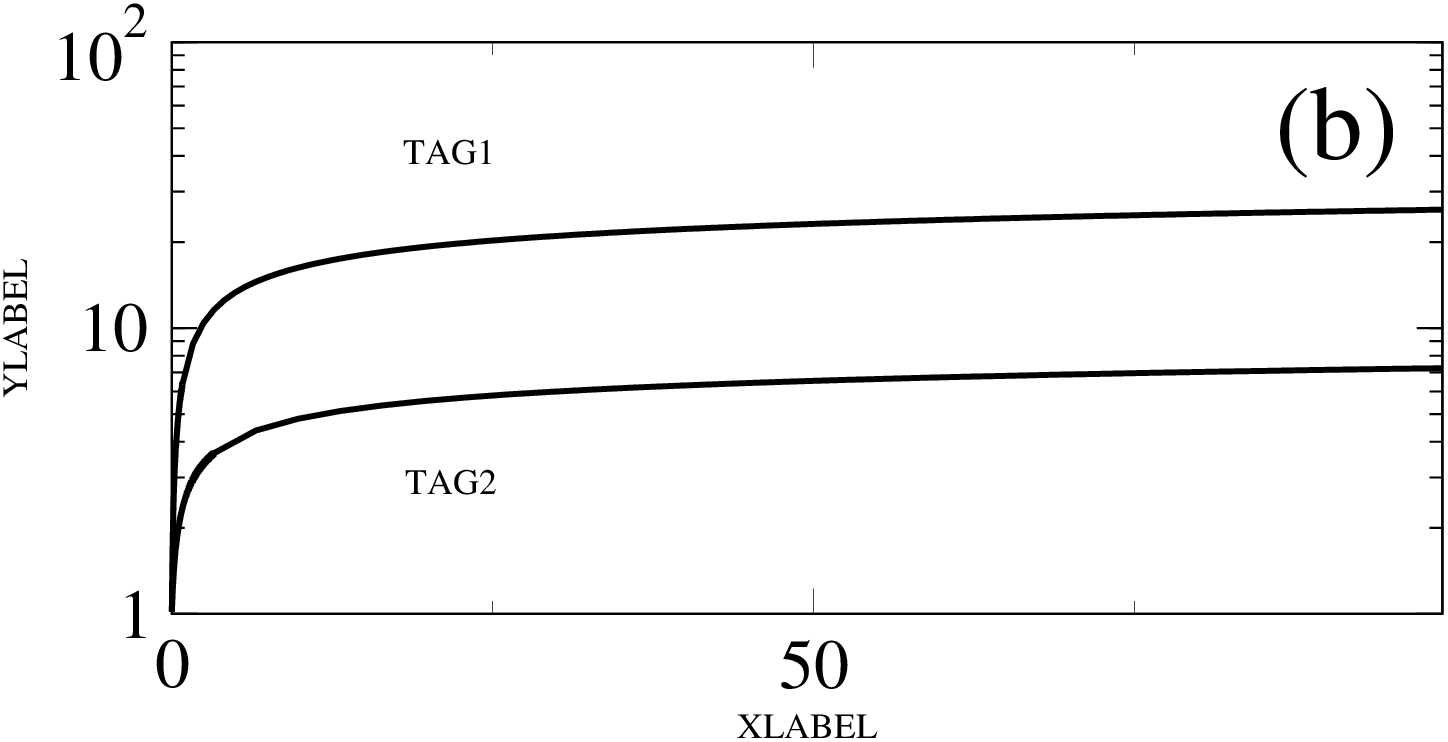}}
\caption{\label{fig4}
Coalescent results for $\langle\sigma^2\rangle/\mu_l$ 
(open symbols)
in comparison with eq. (\ref{eq:hudson1}) (thick lines). Thin lines indicate
90\% confidence intervals.
(a) $\langle\sigma_l^2\rangle/\mu_l$  for $n\!=\!2$ and $\Delta/L =10^{-4}$,
as a function of $\el/L$ for $l/L=10^{-4}$ and  $l/L=10^{-3}\,(\circ)$. 
 Also shown are the
values  of $\sigma_l^2/\mu_l$ 
from         eq.~(\ref{eq:Rzero}) for $l/L=10^{-4}$ ($- - -$).
(b) $\langle\sigma_l^2\rangle/\mu_l$ for $n\!=\!2$ and $\Delta/L = 10^{-4}$.}
\end{figure}

\paragraph{SNP clustering:}
We have determined the fluctuations
of $\sigma_l^2/\mu_l$ using
coalescent simulations for sample size $n=2$, proceeding in five steps:
(1) generate a large number of samples of sequence pairs of length $L$
[the results discussed below correspond to values of $L$ ranging 
between $10^6$ and $1.6\,10^8$bp];
(2) determine the SNPs within each sample (each pair of sequences); 
(3) assign the SNPs to contiguous bins 
as illustrated in Figure~\ref{fig:scales}; 
(4) calculate $\sigma^2_l/\mu_l$ for each sample;
(5) finally, average  over samples.
\\[0.2cm]
In these coalescent simulations, the bin size $l$ was taken
to be much smaller than $L$, corresponding to, say the length of an STS compared
to that of a mouse chromosome.
Figure~\ref{fig4}a shows the results of coalescent
simulations
in comparison with eqs.~(\ref{eq:hudson1}) and (\ref{eq:Rzero}).
In keeping with the above discussion, eq. (\ref{eq:hudson1}) is adequate
when $\el$ is much smaller than $L$. In this regime, the fluctuations
of $\sigma_l^2/\mu_l$ are small (but finite
since $N_{\rm bins} = L/l$ is finite). 
As $\el$ approaches $L$, eqs.~(\ref{eq:hudson1}) and (\ref{eq:Rzero})
are inappropriate, and the fluctuations increase significantly,
as expected.
\\[0.2cm]
Three qualitative observations emerge from our simulations:
\begin{enumerate}
\item in the region $\Delta < \el \ll L$, the observed
values of $\sigma_l^2/\mu_l$ are larger
than unity. If $\el$ is much smaller than $\Delta$, 
or if $\el$ approaches $L$,
 $\sigma_l^2/\mu_l\rightarrow 1$; 
 \item for small
 values of $l$, $\sigma_l^2/\mu_l$ exhibits
 a plateau for intermediate values of $\el$ (indicated
 by a dashed line in Figure~\ref{fig4}a);
 \item  
 for larger values of $l$, the plateau disappears.
 \end{enumerate}
 These qualitative observations can be understood as follows.
 \begin{enumerate}
 \item In the absence of recombination, where  $\el = L$, a uniform SNP
distribution is expected. In this case
\begin{equation}
\sigma_l^2 /\mu_l = 1 
\end{equation}
[contradicted by equation (12)? and if not, why not?] 
as pointed out above.
Conversely, if $\el$ is much smaller than $l$ and $\Delta$, 
the Poisson process averages over the fluctuating local rates
$\lambda(x)$.  One thus expects 
[see eq.~(\ref{eq:inh_poisson})]
local uniformity of SNPs with rate
\begin{equation}
\frac{1}{l}\int_x^{x+l}\!\!{\rm d}x^\prime \lambda(x^\prime)\,.
\end{equation}
Again, $\sigma_l^2/\mu_l = 1$.
In contrast, in the regime  $\Delta < \el \ll L$
significant SNP clustering is observed. 
\item  Consider the situation $l \ll \el \ll  L$. In this regime,
since $l$ is much smaller than $\el$, most bins overlap with only one MRCA segment
and genealogies within a given bin are identical.
Since $\el$ is much smaller than $L$, $ \sigma_l^2/\mu_l$ can
be calculated assuming independent genealogies for each bin.
The result can be obtained from eq.~(11) in the limit
of $C\rightarrow 0$: 
\begin{equation}
\label{eq:Rzero}
\sigma_l^2/\mu_l \simeq 1+l/\Delta\,.
\end{equation}
This means that $\sigma_l^2/\mu_l$ exhibits
a plateau as a function of $\el$  
(its value is equal to $1+l/\Delta$ and thus
independent of $\el$ or $R$). Result (\ref{eq:Rzero}) is
shown as a dashed line in Figure~\ref{fig4}a.
The plateau is cut off 
by  $l$ for small values of $\el$, 
and by $L$ for large values of $\el$. 
\item
There are on average $l/\Delta$ SNPs in
a bin of size $l$. If  $\el$ is much smaller than $l$,  these SNPs
are distributed over many MRCA segments.
If the counts per MRCA segment were statistically independent,
one would expect $\sigma_l^2 \propto \el$, and thus
$\langle \sigma_l^2\rangle$ to increase roughly proportional
to $\el$ (eq. (\ref{eq:hudson1}) shows that
there are logarithmic corrections to this simple model).
As $\el$ approaches $l$, this increase
is cut off; $\langle \sigma_l^2\rangle/\mu_l\rightarrow 1$
as $\el\rightarrow L$. If $l$ is large ($l\stackrel{<}{\sim}L$),
there is no plateau.
\end{enumerate}

\noindent In summary, $\sigma_l^2/\mu_l$  reflects
local SNP clustering. It is expected to be most significant
in the regime $\Delta < \el \ll L$. 
In many organisms $S$ is of the order of $R$.  
For $\!n=\!2$ eq.~(\ref{eq:ell}) implies that $\el$ is roughly $1.5\,\Delta$.
In such cases, recombination alone gives
rise to SNP clustering.
\\[0.2cm]
This clustering is observed
on length scales of the order of $\el$.
Eq. (\ref{eq:hudson1}) 
shows that its 
effect on $\langle \sigma_l^2\rangle/\mu_l$ is
most clearly seen if $l > \el$ (compare Fig.~\ref{fig4}b).
This observation has two
important consequences: (1) in empirical situations
it is advisable to choose the bin size  $l$ 
at least as large as $\el$; (2) deviations from the model considered
may be associated with length scales much longer than $\el$.
Such deviations will be most clearly seen if the 
bin size $l$ is equal to or larger than this length.
 In short: the dependence of $\sigma_l^2/\mu_l$ on $l$
 indicates on which length scales clustering
 of SNPs occurs. 
\\[0.6cm]
\noindent {\sc Data Analysis}
\\[0.2cm]
In the following we discuss the
implications of our analysis  for the interpretation of SNP data from
mouse and human. In both cases we ask whether the neutral model can be rejected.

\begin{figure}
\psfrag{YLABEL}{\large$\sigma_l^2/\mu_l$} 
\psfrag{XLABEL}{\raisebox{-0.0cm}{\large$\el$}}
 \centerline{\includegraphics[clip,width=12cm]{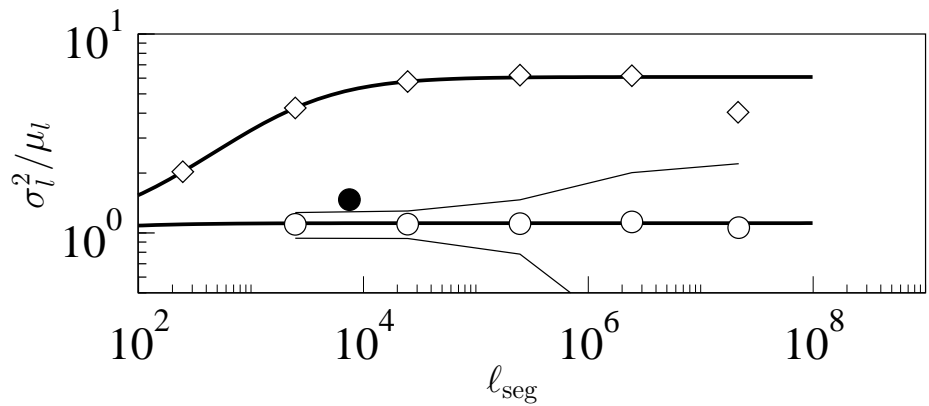}}
 \caption{\label{fig6} 
  $\sigma_l^2/\mu_l$ for {\em M. m. domesticus} SNPs
   \citep{lin00:lar} ($\bullet$). Simulation results for $\langle\sigma_l^2\rangle/\mu_l$
    are shown, corresponding to 
   $l=118\mbox{bp}\, (\circ)$ -- the average read length in
   \citep{lin00:lar} -- and corresponding to  $l=5$kb (Diamond). 
   In addition, the results according to eq. (\ref{eq:hudson1}) (thick
   lines), and, for $l=118\mbox{bp}$,, 90\% confidence intervals (thin lines) are shown.}
\end{figure}

\paragraph{Mouse data:} In their survey of SNPs in mice,
\citet{lin00:lar} observed that SNPs
were not distributed uniformly across the genome. A possible explanation for
this is selective breeding, which has certainly taken place in the
recent evolution of the mice strains investigated. On the other hand, the SNP
clustering in mice might also be due to recombination alone.
\\[0.2cm]
Figure~\ref{fig6} shows the empirically determined value of $\sigma_l^2/\mu_l$
[for {\em M. m. domesticus} SNPs, see \citet{lin00:lar}] in comparison with 
coalescent simulations for $\langle \sigma_l^2\rangle/\mu_l$. 
$L$ was taken to be $1.6\,10^8$bp, corresponding to
the average chromosome length.
A value for $\el$ can be estimated from the average recombination rate
in mice, approximately $0.5$ cM/Mb \citep{nac92:het}.
Here we assume an effective population size $N_{\rm e} = 10000$.
The average bin length $l$ is the average length of the
sequence tagged sites investigated, i.e. $118$bp, smaller than $\Delta$.  SNP
clustering on the scale of $l$ is thus expected to be small. Moreover,
since $l$ is much smaller than $\el$,
$\langle \sigma_l^2\rangle/\mu_l$ is expected to exhibit a plateau as a function of
$\el$, at $\langle \sigma_l^2\rangle/\mu_l = 1.12$. While the plateau is found in
the simulations (Figure~\ref{fig6}), the empirically determined
value of
$\sigma_l^2/\mu_l$ deviates significantly from neutral expectations 
(Figure~\ref{fig6}; $\sigma_l^2/\mu_l = 1.47$). This increase of $\sigma_l^2/\mu_l$
above the value expected under neutrality is consistent with the earlier conclusion that
selection plays
an important role in shaping the genome-wide distribution of
polymorphisms in mice
\citep{lin00:lar}. In order to demonstrate this more conclusively,
long-range
SNP data would be of great interest for two reasons: (1) 
the larger $l$, the larger deviations
from Poisson statistics are expected (ideally,
$l$ would be of the order of $\el$).
Figure~\ref{fig6} shows the
increase
in neutral SNP clustering if $l$ is increased to 5kb. (2) Selection may
act on length scales
greater than $\el$ and may thus
contribute only
weakly at very small values of $l$ such as those corresponding to an
average
STS.

\begin{figure}
\psfrag{YLABEL}{\large$\sigma_l^2/\mu_l$} 
\psfrag{XLABEL}{\raisebox{-0.0cm}{\large$\el$}}
\psfrag{YLABEL1}{\large$P(n_j(l)\!=\!k)$} 
\psfrag{XLABEL1}{\raisebox{-0.0cm}{\large$k$}}
\psfrag{TAG2}{\raisebox{-0.0cm}{\large$l=200$kb}}
\psfrag{TAG1}{\raisebox{-0.0cm}{\large$l=460$bp}}
\psfrag{TAG3}{\raisebox{-0.0cm}{(1)}}
\psfrag{TAG4}{\raisebox{-0.0cm}{(2)}}
\psfrag{CUTOFF}{\raisebox{-0.0cm}{\large cut-off}}
 \centerline{\includegraphics[clip,width=12cm]{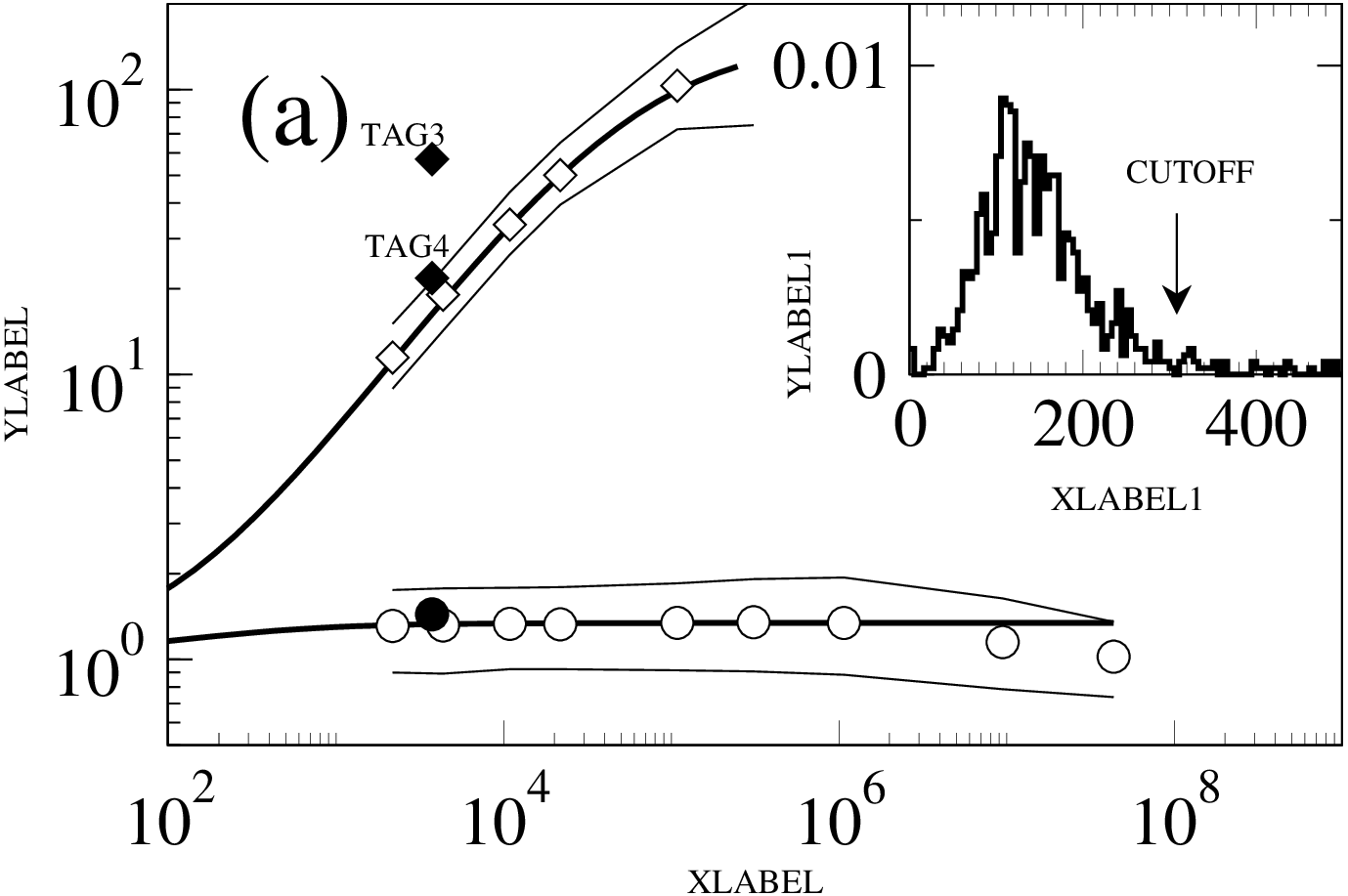}}
\mbox{}\\[0.2cm]
\psfrag{YLABEL}{\large$\sigma_l^2/\mu_l$} 
\psfrag{XLABEL}{\raisebox{-0.0cm}{\large$\el$}}
 \centerline{\includegraphics[clip,width=12cm]{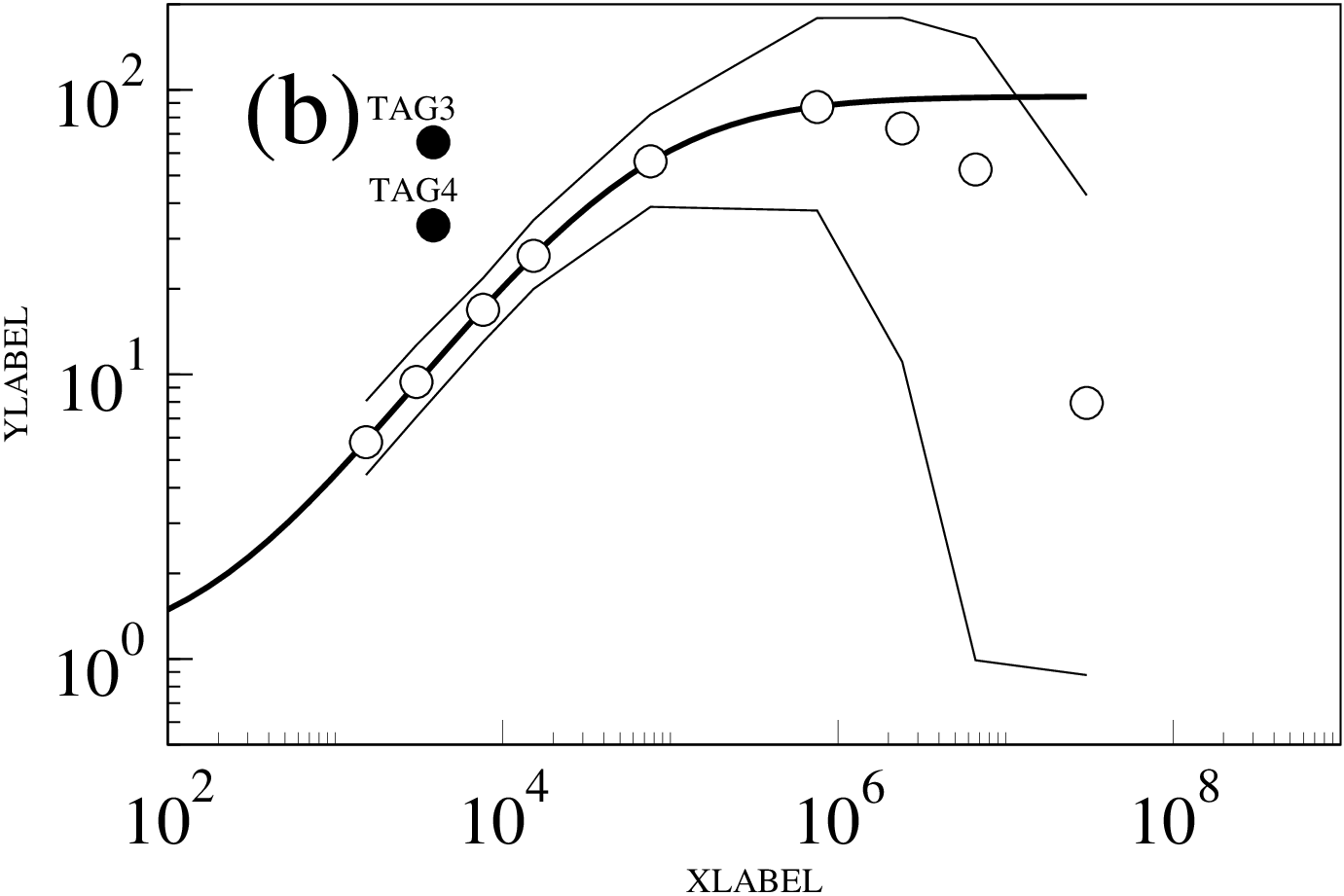}} 
 \caption{\label{fig7} 
  (a) Variance of SNPs for chromosome 6 in the human genome. Empirical data
  for bin sizes $l=460\mbox{bp}\, (\bullet$) and $l=200\mbox{kb}\,(\blacklozenge)$ 
  are determined
  from the data provided by \citet{tsc01:map}. 
   The data points
   labeled by $(1)$ and $(2)$ differ by a choice of cut-off (see text).
  Also shown are the mean results of coalescent simulations corresponding to  $l=460\mbox{bp}\,(\circ)$ and
  $l=200\mbox{kb}$ (Diamond) and their 90\% confidence intervals (thin lines),
compared to theoretical expectations from (\ref{eq:hudson1}) (thick
  lines).
  The inset shows the empirical
  distribution of $n_j(l)$ corresponding to $l=200$kb.
  (b) Variance of SNPs for the human  X chromosome. Empirical data
   for $l=200$kb $(\bullet)$ were determined from the data provided by \citet{tsc01:map}.
  Also shown are mean results of coalescent simulations corresponding to
$l=200\mbox{kb}\,(\circ)$, and their 90\% confidence intervals (thin lines)
compared to theoretical expectations from equation (\ref{eq:hudson1}) (thick line).
   }
\end{figure}

\paragraph{Human chromosome 6:} The distribution of chromosome-wide human
SNP data was      
empirically determined for $l=460$bp and $l=200$kb by \citet{tsc01:map}. 
Figure~\ref{fig7}a shows our coalescent simulations
compared
to the empirical data for these length scales. Consider first the case
of
$l=460$bp. In the simulations,
$\langle\sigma_l^2\rangle/\mu_l$ exhibits a plateau for
intermediate values of $\el$ at $\langle\sigma_l^2\rangle/\mu_l\simeq 1.34$
[according to eq.~(\ref{eq:Rzero})].
This implies that the choice of $\el$ attributed to the empirical
value is uncritical.
Empirically, $\sigma_l^2/\mu_l = 1.44$. 
From Figure~\ref{fig7}a we conclude:
given the degree SNP clustering observed in the human genome on scales
of the
order of $l=460$bp, the neutral model cannot be rejected with confidence.
\\[0.2cm]
The
situation for $l=200$kb is very different. In this case, the
simulation
results do not exhibit a plateau. The numerical results indicate that
$\langle\sigma_l^2\rangle/\mu_l$ increases roughly with $\el$ (for $\el \ll
l$), as
suggested  above. Furthermore, the empirical value for $\sigma_l^2/\mu_l$
(labeled
(1) in Figure~\ref{fig7}a) lies significantly above the values for the
neutral
infinite-sites model with recombination [the corresponding
value of $\el$ was estimated assuming an effective population size
of $N_{\rm e} = 10000$ and a recombination rate of $1$ cM/Mb \citep{pri01:lin}]. 
This deviation is possibly
caused by
selection:  the HLA system, which contains more than 100 genes and spans more than 4 Mb
on the short arm of  chromosome 6 \citep{kle93:mhc}, 
has an exceptionally high SNP density. This is maintained by balancing
selection
\citep{tsc01:map,ohu00:imp}. The inset of Figure~\ref{fig7}a shows the
empirically
determined distribution of the number of SNPs per bin $P(n(l)=k)$. It
exhibits a
strong tail for large values of $k$, which may be due to selection. By
ignoring
this tail the estimate (labeled (2) in Figure~\ref{fig7}a) for
$\sigma_l^2/\mu_l$ is
reduced considerably. Given the uncertainty as to which value of $\el$
should be
assigned to the data points, one may argue that this second estimate is
consistent with the neutral infinite-sites model.

\paragraph{Human X chromosome:} Finally, Figure~\ref{fig7}b shows the
empirical
estimate of $\sigma_l^2/\mu_l$ given $l=200$kb for
the X chromosome. This empirical estimate was reduced by discarding the tail in the
distribution $P(n(l)=k)$ for large $k$ as was done in the analysis of the
chromosome 6 data. However, both estimates of SNP clustering on  the human X
chromosome deviated significantly from neutrality.
This observation is consistent
with the fact that due to its hemizygosity in males chromosome X 
should be
affected more by selection than autosomes such as chromosome 6.
\\[0.6cm]
\noindent {\sc Discussion}
\\[0.2cm]
In this paper we investigate whether
the observations of SNP clustering in mice and humans are compatible with neutral
expectations. We chose the variance in the number of SNPs found in equal-length contiguous
divided by the mean number of SNPs found in each bin as a measure of SNP
clustering: $\sigma_l^2/\mu_l$. Since under a Poisson distribution the variance is
equal to the mean, $\sigma_l^2/\mu_l=1$ in the absence of recombination. If SNPs
are clustered, $\sigma^2_l/\mu_l>1$. 
\\[0.2cm]
Whether or not SNP clustering is significant depends
on the relative sizes of the mean spacing
between neighboring SNPs, $\Delta$, the mean length of segments with constant
time to the most recent common ancestor, $\el$, and the total length of the
chromosome, $L$. 
Specifically,
clustering is observed if $\Delta < \el\ll L$. In contrast, if recombination is
either very frequent compared to mutation ($\el\ll\Delta$) or very rare ($\el\ll L$),
no clustering is observed\footnote{Rather than the
recombination rate $R$ we use the corresponding length
scale $\el$ as our point of reference
for discussing SNP clustering; $\el$ can directly be compared to the
other length scales of the problem, viz. the length of the
bins in which SNPs are sampled in experiments, $l$, as well as $\Delta$, and $L$. Moreover,
equation (\ref{eq:ell}) shows that $\el$ is a simple function of $L$, $R$, and
the sample size $n$, thereby establishing the link between simulations
conditioned on $R$ and our observations.}.%
\\[0.2cm]
We have shown that it is essential to 
consider the effect of the scale on which SNPs are
sampled, $l$. In our simulations $l$ describes the length of contiguous
non-overlapping bins. This length is short compared to the length of the
chromosome, and the corresponding large number of such bins leads to narrow
confidence intervals around $\sigma^2_l$ for the biologically relevant
parameters. As a result, meaningful comparisons between model and observation
can be made. In the case of the mouse, the neutral model is rejected with
marginal significance.
This conclusion depends on the assumption of a ``true'' recombination rate for
mice. This is difficult to know, but Figure (\ref{fig6}) shows that our
hypothesis test is quite robust with respect to errors in the estimation of $R$
(and hence $\el$).
\\[0.2cm]
In the case of human chromosome 6, the influence of $l$ on the outcome of the
neutrality test was striking. Significant SNP clustering was observed for large
but not for small $l$ (Figure \ref{fig7}a). For large bins ($l=200$ kb) the
distribution of the number of SNPs per bin had a strong positive skew (Figure
\ref{fig7}a, inset). By cutting off the tail of bins containing many SNPs,
clustering was reduced to its neutral level.
\\[0.2cm]
No such effect of cutting off the tail of SNP-dense bins was observed for the
X-chromosome. It therefore constitutes the most significantly non-neutral SNP
collection among the three data sets investigated in this study.
\\[0.2cm]
Factors that might lead to such a rejection of the neutral model include
population expansion, population subdividion, and variation in (physical)
mutation rate.  In the case of the data sets we have investigated, some
illuminating biology pertaining to these factors is known. \citet{lin00:lar}
tested whether unequal physical mutation rates could account for their
observation of SNP clustering in mice. Their approach was to resequence 16 STSs
with no SNPs and 16 STSs with five or more SNPs from closely related species of
mice. They observed that the classification of high-scoring and low-scoring STSs
was not reproduced in these other taxa and concluded from this that fluctuations
in inherent mutation rates could not account for the observation of
significant SNP clustering. The claim that selective breeding has been
important in shaping the SNP distribution in mice is plausible, but other factors such
as population expansion and subdivision can presumably not be ruled out.
\\[0.2cm]
The situation is slightly different in the case of human chromosomes 6 and X,
where deviation from the neutral model was much more pronounced for the sex
chromosome than for the autosome. Since all chromosomes have undergone the same
history of population expansion and migration, selection seems to be the only
explanation for this difference. The hemizygosity in males of the X-chromosome,
which makes most deleterious mutations dominant in males, fits well with this
conclusion. 
\\[0.6cm]
\noindent {\sc Acknowledgements}
We would like to thank Richard Hudson for discussion and much appreciated
comments on an earlier version of this manuscript. This work was supported by a
grant from the Swedish Science Foundation.

\bibliography{../references}
\newpage

\end{document}